\documentclass[superscriptaddress,twocolumn,floatfix,aps]{revtex4-2}
\usepackage[utf8]{inputenc}
\usepackage{xcolor}
\usepackage{booktabs}
\usepackage{bbm}
\usepackage{bm}
\usepackage{physics}
\usepackage{graphicx}
\usepackage{placeins}
\usepackage{float}
\usepackage{chronology}
\usepackage{amsmath,amsfonts,amssymb,amsthm}
\usepackage{qcircuit}
\usepackage{indentfirst}
\usepackage{dsfont}
\usepackage[normalem]{ulem}
\usepackage{mathtools}
\usepackage{hyperref}
\usepackage{tcolorbox}
\usepackage{array}
\usepackage{multirow}

\begin{document}

\title{Quantum readout error mitigation via deep learning}
\author{Jihye Kim}
\affiliation{SKKU Advanced Institute of Nanotechnology and Department of Nano Engineering, Suwon, 16419, Republic of Korea}
\author{Byungdu Oh}
\affiliation{SKKU Advanced Institute of Nanotechnology and Department of Nano Engineering, Suwon, 16419, Republic of Korea}
\author{Yonuk Chong}
\affiliation{SKKU Advanced Institute of Nanotechnology and Department of Nano Engineering, Suwon, 16419, Republic of Korea}
\author{Euyheon Hwang}
\email{euyheon@skku.edu}
\affiliation{SKKU Advanced Institute of Nanotechnology and Department of Nano Engineering, Suwon, 16419, Republic of Korea}
\author{Daniel K. Park}
\email{dkp.quantum@skku.edu}
\affiliation{SKKU Advanced Institute of Nanotechnology and Department of Nano Engineering, Suwon, 16419, Republic of Korea}

\begin{abstract}
Quantum computing devices are inevitably subject to errors. To leverage quantum technologies for computational benefits in practical applications, quantum algorithms and protocols must be implemented reliably under noise and imperfections. Since noise and imperfections limit the size of quantum circuits that can be realized on a quantum device, developing quantum error mitigation techniques that do not require extra qubits and gates is of critical importance. In this work, we present a deep learning-based protocol for reducing readout errors on quantum hardware. Our technique is based on training an artificial neural network with the measurement results obtained from experiments with simple quantum circuits consisting of singe-qubit gates only. With the neural network and deep learning, non-linear noise can be corrected, which is not possible with the existing linear inversion methods. The advantage of our method against the existing methods is demonstrated through quantum readout error mitigation experiments performed on IBM five-qubit quantum devices.
\end{abstract}

\maketitle

\section{Introduction}

The theory of quantum computation opens up tremendous opportunities as it promises drastic computational benefits for solving commercially relevant problems~\cite{10.2307/2899535,zalka1998simulating,shor1999polynomial,PhysRevLett.103.150502_HHL_qBLAS,Preskill2018quantumcomputing}. One of the major technological hurdles against practical quantum advantage is the unavoidable noise and imperfections. Although the theory of quantum error correction and fault-tolerance guarantees that imperfections are not the fundamental objections to quantum computation ~\cite{shor1995scheme,steane1996error,calderbank1996good,fowler2012surface,RevModPhys.87.307}, the size of quantum circuits needed to realize the theory is beyond the reach of the near-term technology. Consequently, in near-term quantum computing, all physical qubits are expected to be operating as logical qubits, leading to the Noisy Intermediate-Scale Quantum (NISQ) era~\cite{Preskill2018quantumcomputing}. Therefore, to bridge the gap between theoretical results and experimental capabilities, developing algorithmic means at the software level to reduce quantum computational errors without increasing the quantum resource overhead, such as the number of qubits and gates, is desired~\cite{PhysRevLett.117.260501,Songeaaw5686,PhysRevX.8.031027,PhysRevLett.119.180509,9226505}.

This work addresses the problem of reducing the readout errors that occur during the final step of the quantum computation. We propose a machine learning-based protocol to reduce the quantum readout error. Our method trains an artificial neural network with real measurement outcomes from quantum circuits with known final states as input and ideal measurement outcomes of the same circuit as output. After training, the neural network is used to infer the ideal measurement outcomes from real measurement outcomes of arbitrary quantum computation. Since our quantum readout error mitigation (QREM) protocol only relies on training a classical neural network with known measurement outcomes, the extra quantum resource arises only for gathering noisy measurement results. This is done with a random state prepared by applying an arbitrary single-qubit gate to the default initial state of qubits that are subject to measurement in a quantum algorithm. Thus our method does not increase the number of qubits and gates beyond what is needed by the quantum algorithm itself.

Using the neural network and deep learning, non-linear readout errors can be corrected. This feature places our QREM method above the predominantly used methods based on linear inversion that do not take non-linear effects into account. Moreover, while the linear inversion method often produces non-physical results and requires post-optimization, our method always produces physically valid results. We compare our QREM with the linear inversion method through experiments with superconducting quantum devices available on the IBM cloud platform for two to five qubits. The experimental results demonstrate that the amount of readout error reduced by our method exceeds that of the previous method in all instances differentiated by the number of qubits and the error quantifier selected from mean squared error, Kullback-Leibler divergence and infidelity.

The remainder of the paper is organized as follows. Section~\ref{sec:framework} describes the theoretical framework of this work, such as the quantum readout error, existing mitigation methods for it, and the underlying idea of our neural network-based approach to the problem. Section~\ref{sec:ML} explains the machine learning algorithm used in this work for the quantum readout error mitigation. Section~\ref{sec:exp_results} presents the proof-of-principle experiment performed with three five-qubit quantum computers available on the IBM Quantum cloud platform. Conclusions and directions for future work are provided in Sec.~\ref{sec:conclusion}.

\section{Theoretical Framework}
\label{sec:framework}
Quantum computation can be thought of as a three-step procedure consisting of input state initialization, unitary transformation, and readout. In experiments, the readout step is usually implemented as the projective measurement in the computational basis, which collapses an $n$-qubit final state $|\psi_j\rangle\in\mathbb{C}^{2^n}$ to $|i\rangle$ with the probability $p(i|j) = \vert\braket{i}{\psi_j}\vert^2$. Without loss of generality, we denote $p(i) \coloneqq p(i|j)$. In the ideal (error free) case, the final result of a quantum algorithm is determined by the probability distribution of the measurement outcomes, denoted by
$$\bm{p} = \lbrace p(0), p(1), \ldots, p(2^n-1) \rbrace.$$
Equivalently, $\bm{p}$ is the set of diagonal elements of the final density matrix $\rho$ produced at the end of the quantum computation. Many quantum algorithms are designed in a way that the solution to the problem is encoded in the probability distribution $\bm{p}$. In particular, estimating an expectation value of some observable from such probability distribution is central to many NISQ algorithms, such as Variational Quantum Eigensolver (VQE)~\cite{peruzzo_variational_2014,McClean_2016,kandala_hardware-efficient_2017,barron2020measurement}, Quantum Approximate Optimizatoin Algorithm (QAOA)~\cite{farhi2014quantum}, Quantum Machine Learning (QML)~\cite{Havlicek2019,blank_quantum_2020,PARK2020126422}, and simulation of stochastic processes~\cite{blank_quantum-enhanced_2021,ho2021enhancing}.

\begin{figure}[t]
    \centering
    \includegraphics[width=0.85\columnwidth]{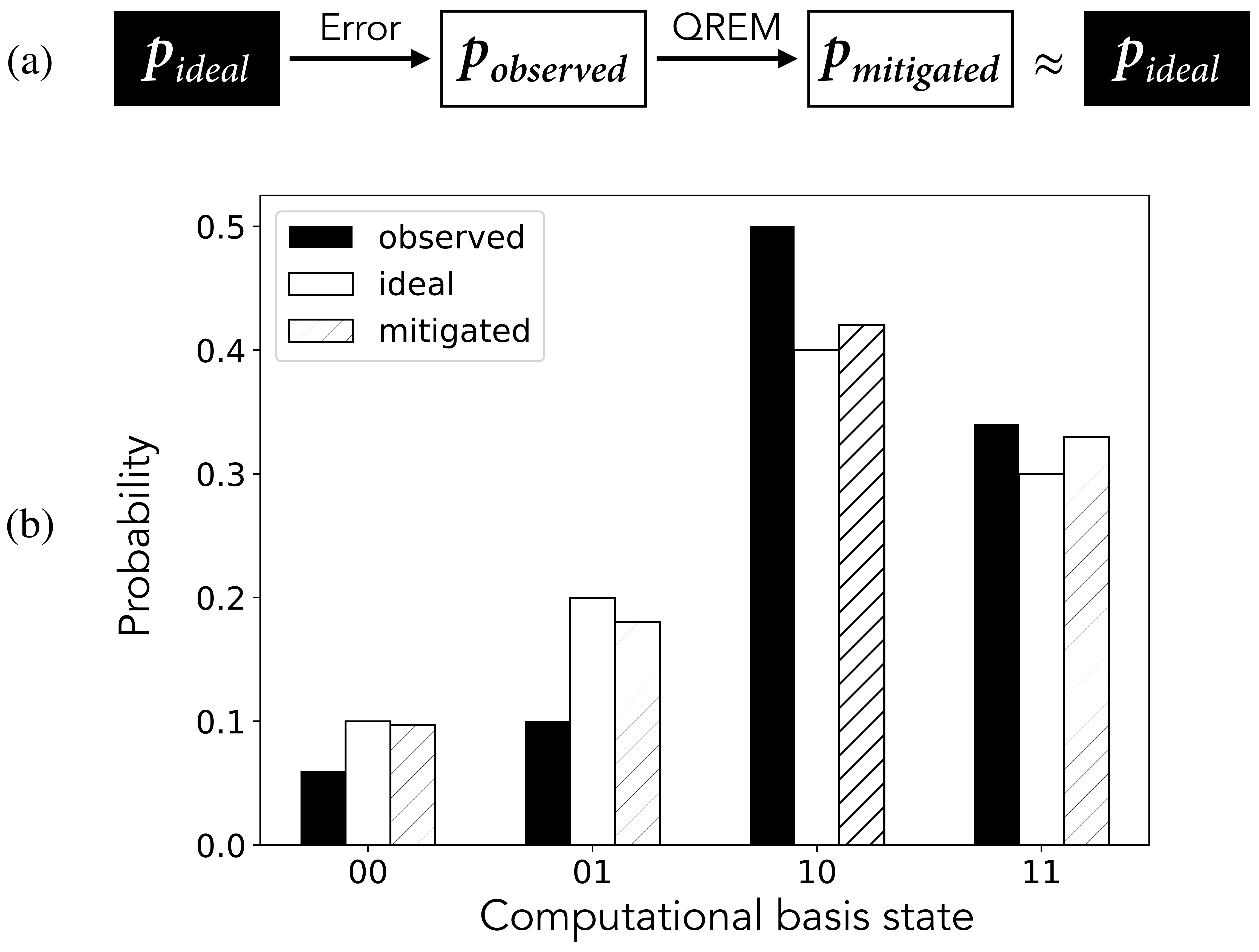}
    \caption{\label{fig:1}An illustration of the quantum readout error mitigation (QREM) procedure. In the noiseless case, the underlying quantum algorithm should generate a probability distribution $\boldsymbol{p}_{ideal}$ upon the measurement in the computational basis. However, due to imperfections and noise, one obtains the noisy distribution $\boldsymbol{p}_{observed}$, which deviates from the true distribution. The QREM method constructs a probability distribution $\boldsymbol{p}_{mitigated}$ based on $\boldsymbol{p}_{observed}$ that is as close to $\boldsymbol{p}_{ideal}$ as possible. The bar graph in (b) is an artificially generated two-qubit example that illustrates the role of QREM. After a QREM algorithm, $\boldsymbol{p}_{mitigated}$ is a more accurate description of the true probability distribution than the raw data.}
\end{figure}

However, due to readout errors, the observed probability distribution can deviate from the ideal probability distribution. Hereinafter, we denote the probability to observe $|i\rangle$ in the experiment as $\hat{p}(i)$ and the error map as $\mathcal{N}$ that transforms the ideal probability vector to an observed one as $\hat{\bm{p}} = \mathcal{N}(\bm{p})$. Since $\mathcal{N}$ describes the transformation of a probability distribution, it must preserve the 1-norm.
The goal of QREM is to minimize the loss function
\begin{equation}
\label{eq:loss_function}
    D\left(\bm{p},\mathcal{N}(\bm{p})\right),
\end{equation}
defined by some distance measure $D$ for probability distributions. The goal of QREM is conceptually depicted at the top of Fig.~\ref{fig:1}, with an example of a two-qubit case given in the bottom.

Since QREM is of critical importance especially for NISQ computing, several protocols have been proposed recently to address this issue. In essence, these methods assume that the noisy function is linear and relies on solving the linear equation
\begin{equation}
\label{eq:linear_model}
    \hat{\bm{p}} = \Lambda \bm{p},
\end{equation}
where the linear response matrix $\Lambda$ is estimated by some tomographic means~\cite{kandala_hardware-efficient_2017,Qiskit-Textbook,PhysRevA.100.052315,Maciejewski2020mitigationofreadout,nachman_unfolding_2020,9259938,PhysRevA.103.042605,smith2021qubit}. This usually demands $O(2^n)$ measurements with an assumption that the computational basis state preparation error is negligible compared to the readout error. The number of experiments can be reduced with certain assumptions about the noise model, such as those dominated by few-qubit correlations~\cite{geller2020efficient,9259938,PhysRevA.103.042605} and by independent single-qubit Pauli errors acting on the final state during the finite time between the last gate and the detection event~\cite{9142431}. After $\Lambda$ is found, correct measurement outcomes are predicted by calculating $\Lambda^{-1}\hat{\bm{p}}$. We refer to this technique as linear inversion (LI) based QREM (LI-QREM). This approach often requires extra classical post-processing~\cite{Maciejewski2020mitigationofreadout,nachman_unfolding_2020,PhysRevA.103.042605}, since the inversion may produce a vector that is not a probability distribution. In general, the readout noise is not static. Hence the QREM procedure described above needs to be frequently implemented as a part of the calibration routine of the given experimental setup.

\begin{figure}[t]
    \centering
    \includegraphics[width=1\columnwidth]{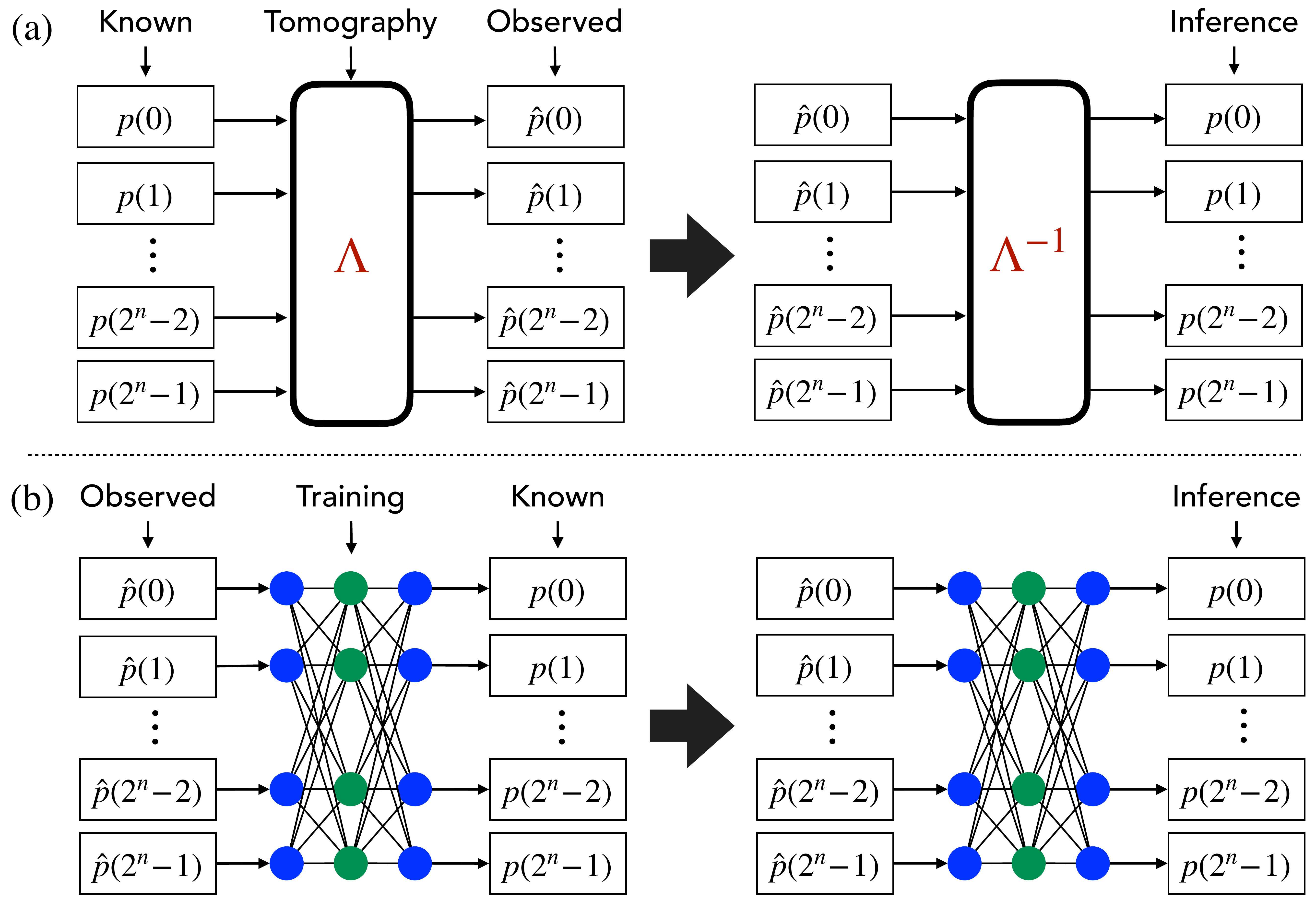}
    \caption{QREM based on (a) linear error model and readout tomography (LI-QREM) (b) deep learning (NN-QREM). The deep learning model is trained to implement the function $\mathcal{F}$, which serves as an approximation to the inverse of the true error map, i.e. $\mathcal{F}\approx\mathcal{N}^{-1}$.\label{fig:2}}
\end{figure}

While this work shares the same goal as the previous methods, we leverage classical deep learning techniques to approximate $\mathcal{N}^{-1}(\bm{p})$. Namely, we train a neural network (NN) denoted by $\mathcal{F}$ which describes the map $\bm{p} = \mathcal{F}(\hat{\bm{p}})$ such that $\mathcal{F}\approx \mathcal{N}^{-1}$. Since the deep learning model can take care of spurious non-linear effects, the amount of error suppression is expected to be beyond what the linear model can achieve. Moreover, by using the softmax function in the final layer of a neural network, one can ensure that the output always represents a probability distribution. Figure~\ref{fig:2} compares the basic idea of previous LI-QREM and the neural network based method proposed in this work, which we refer to as NN-QREM.
The following section explains the machine learning model in detail.

\section{The Machine Learning Algorithm}
\label{sec:ML}

\subsection{Data collection for training}

\begin{figure}[h]
    \centering
    \includegraphics[width=0.75\columnwidth]{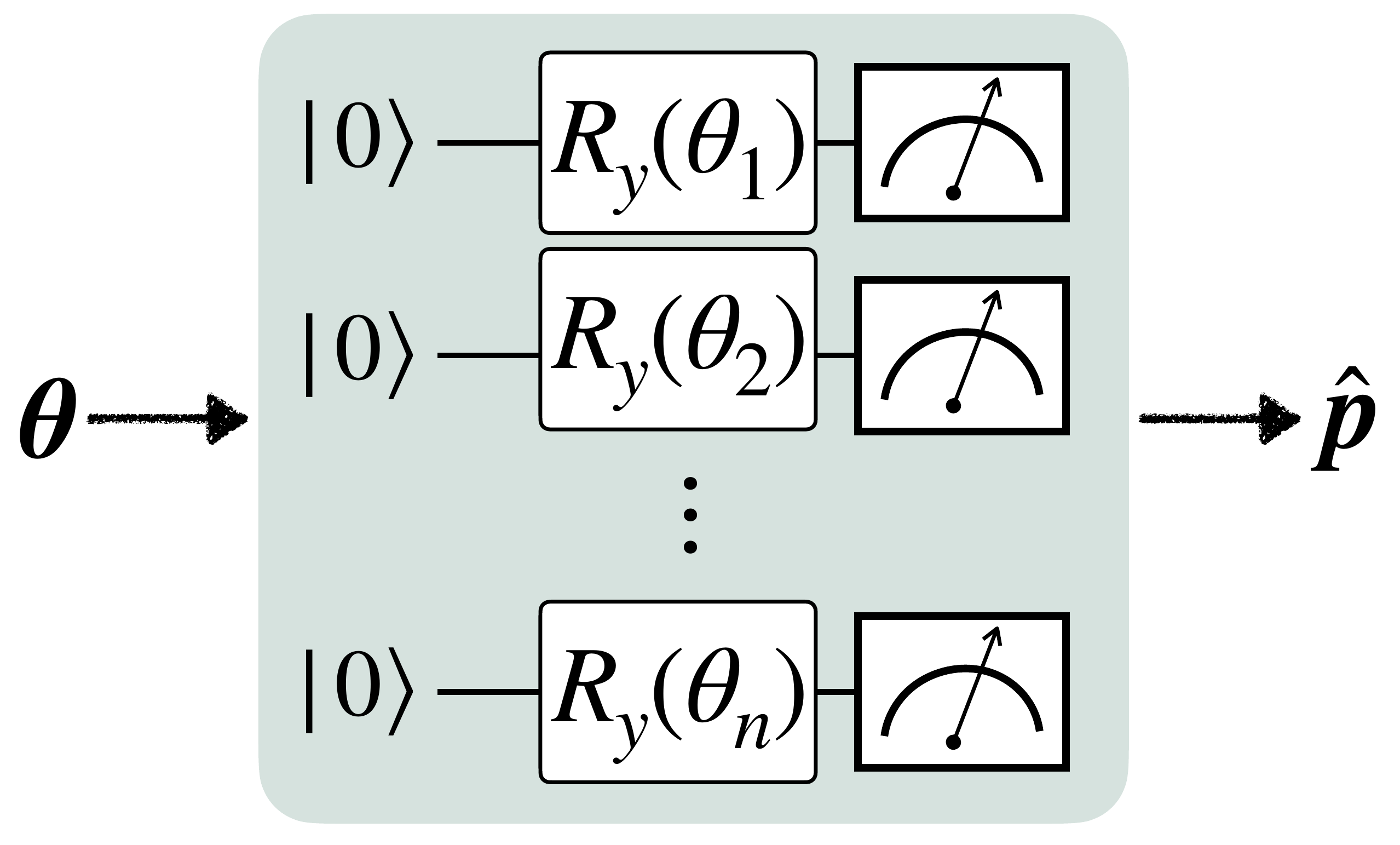}
    \caption{\label{fig:3}The quantum circuit for generating the noisy probability distribution $\hat{\bm{p}}$ used as an input to the neural network during training. The training data is generated by applying an arbitrary single-qubit rotation $R_y(\theta)$ to each qubit and measuring qubits in the computational basis. Training of the neural network also requires the ideal probability distribution $\bm{p}$, and this is calculated using the set of rotation angles $\bm{\theta}$ used in the quantum circuit.}
\end{figure}

In order to train a deep neural network for QREM, we need to have both $\bm{p}$ and $\hat{\bm{p}}$, meaning that qubits need to be prepared in some known state before measurement. Since single-qubit gate errors are usually negligible when compared to those of two-qubit gates and measurement in modern quantum devices~\cite{PhysRevA.100.052315,Maciejewski2020mitigationofreadout,https://doi.org/10.1002/qute.202000005}, we prepare the training set of quantum states using single-qubit gates only. In particular, the training set is constructed by applying $R_y(\theta)$ gate, which corresponds to the rotation around the y-axis of the Bloch sphere, to all qubits in the system with randomly and independently generated angle $\theta\in \lbrack 0,2\pi)$. Since the measurement is performed in the computational basis, which is the $\sigma_z$ basis by convention, there is no need to apply the $R_z$ gate. Hence the quantum circuit depth for generating a training data is one. Figure~\ref{fig:3} represents the quantum circuit that generates $\hat{\bm{p}}$ as an input to the neural network for training. The ideal probability distribution $\bm{p}$, which is inserted as the output of the neural network during training, can easily be calculated from the rotation angles. For an $n$-qubit system, the probability to measure a computational basis state $b\in\lbrace 0,1\rbrace^n$ (i.e. an $n$-bit string) is
$$
p(b) = \left\vert\prod_i^{n} \cos^{1-b_i}(\theta_i/2)\sin^{b_i}(\theta_i/2)\right\vert^2,
$$
where $b_i$ is the $i$th bit of the binary string $b$.

\subsection{Model construction} 
The deep learning model is constructed with an input layer, hidden layers, and an output layer (see Fig.~\ref{fig:2} (b)). Input layer and output layer have $2^{n}$ nodes, whose numerical values represent the probability of measuring the computational basis states in the actual experiment and the ideal case, respectively. All hidden layers are fully connected layers, and each hidden node employs the Rectified Linear Unit (ReLU) as the activation function. The output layer uses the softmax function for activation, which ensures that the output represents a probability distribution. The loss function for optimizing the weights and biases of the neural network is the categorical cross entropy, which is normally used in multi-label classification problems. The free parameters are updated by the Adam optimizer \cite{kingma2014adam} whose hyperparameters, such as the learning rate, are empirically chosen (see Sec.~\ref{sec:exp_results}).

\subsection{Inference}
The trained neural network represents the function $\mathcal{F}\approx\mathcal{N}^{-1}$. Thus, the error-mitigated probability distribution, which is denoted by $\tilde{\bm{p}}$, is obtained by inserting $\hat{\bm{p}}$ from an experiment of interest as the input to the trained neural network. The inference can be expressed as $\tilde{\bm{p}} = \mathcal{F}(\hat{\bm{p}})\approx \mathcal{N}^{-1}(\hat{\bm{p}})$.

\section{Experiment}

\subsection{Experimental Setup}
\label{sec:exp_results}

Two quantum readout error mitigation techniques, namely LI-QREM and NN-QREM were tested on three different five-qubit quantum computers available on IBM Quantum Experience. Training of the neural network and inference are carried out as described in Sec.~\ref{sec:ML} using Keras library of Python. The LI-QREM results are obtained by using the default readout error mitigation package in Qiskit Ignis~\cite{Qiskit} (i.e. $\mathtt{CompleteMeasFitter}$). We use the least squares method to calibrate the resulting matrix from the Qiskit library to ensure physical results. 

\begin{figure}[h]
    \centering
    \includegraphics[width=0.9\columnwidth]{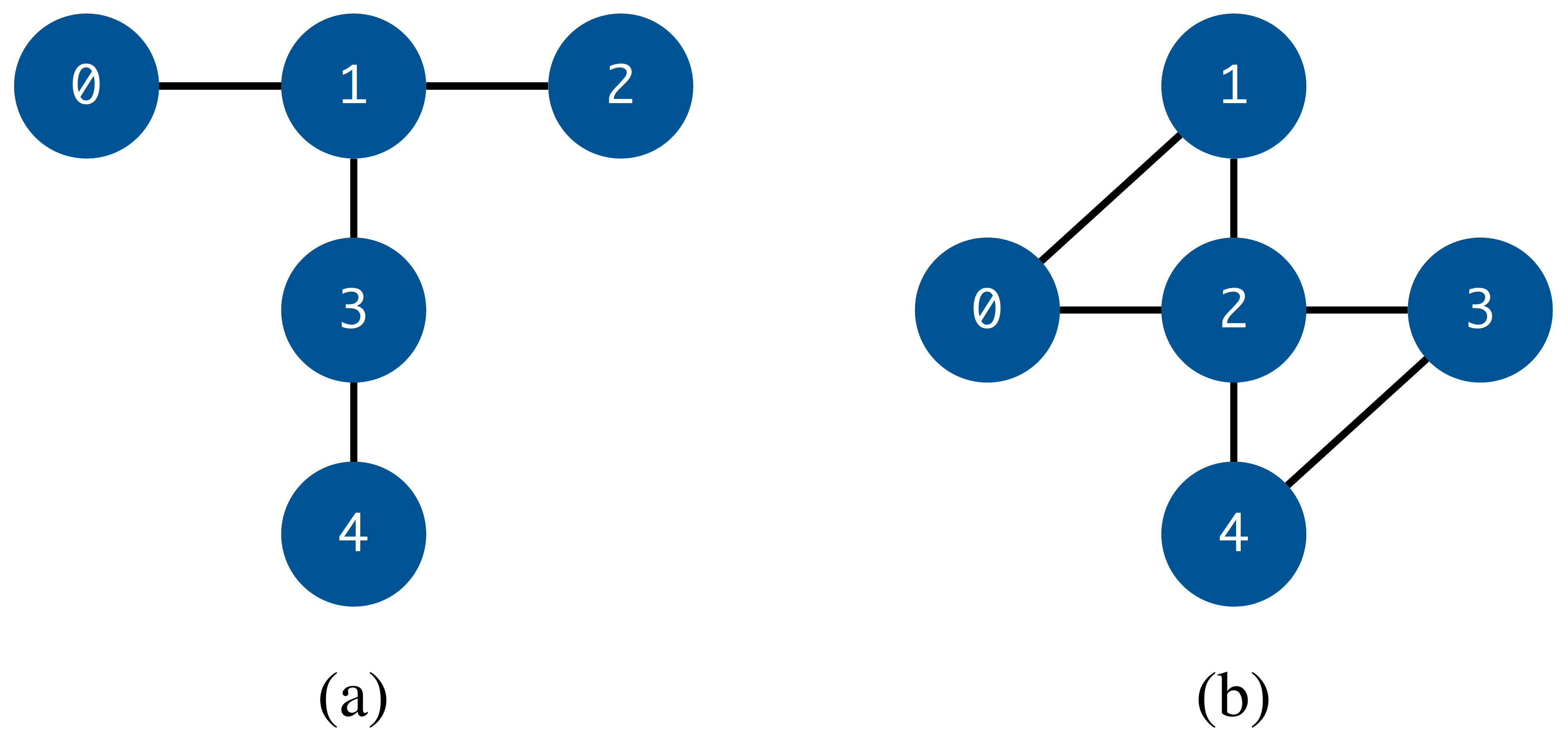}
    \caption{\label{fig:devices} Coupling map (qubit connectivity) of the quantum devices used in this work. \texttt{ibmq\_quito} and \texttt{ibmq\_belem} have the arrangement shown in (a), and \texttt{ibmq\_qx2} has the arrangement shown in (b).} 
\end{figure}

On the five-qubit devices, QREM of two to five qubits are carried out. The quantum devices with smaller quantum volume were chosen in this study since the error mitigation is more critical for such devices. Furthermore, we chose devices of different qubit connectivity as shown in Fig.~\ref{fig:devices}. For the first arrangement (Fig.~\ref{fig:devices} (a)), two quantum devices were selected based on the amount of queue on the cloud service, namely \texttt{ibmq\_quito} and \texttt{ibmq\_belem}, to minimize the experimental time. The two- and three-qubit QREM were performed on \texttt{ibmq\_quito}, and the four- and five-qubit QREM experiments were performed on \texttt{ibmq\_belem}. For the second arrangement, \texttt{ibmq\_qx2} was the only device available for us to use. Thus for this device, all two- to five-qubit experiments were implemented.

Several hyperparameters need to be fixed when training a neural network. In our experiments, we set the number of nodes in each hidden layer to be $5 \times 2^n$ to ensure that the number of nodes scales only linearly with the size of the probability distribution. The learning rate for the Adam optimization algorithm was fine-tuned for each number of qubits. The number of hidden layers was optimized using 5-fold cross-validation (see Supplementary Information). The hyperparameter values of the neural networks used in the NN-QREM experiments are provided in Tab.~\ref{table:1} and Tab.~\ref{table:2} for the two device types shown in Fig.~\ref{fig:devices}, respectively.

\begin{table}[ht]
\begin{tabular}{@{}>{\centering\arraybackslash}m{6.5em}|
>{\centering\arraybackslash}m{3.5em}
>{\centering\arraybackslash}m{3.5em}
>{\centering\arraybackslash}m{4.5em}
>{\centering\arraybackslash}m{4.5em}@{}}
\toprule
\begin{tabular}[c]{@{}c@{}}Num. of\\ qubits\end{tabular}                       & 2     & 3     & 4                 & 5                 \\ \midrule
\begin{tabular}[c]{@{}c@{}}Num. of\\ training data\end{tabular}                & 1175  & 3472  & 9700              & 9700              \\
\midrule
\begin{tabular}[c]{@{}c@{}}Num. of\\ test data\end{tabular}                    & 200   & 200   & 200               & 200               \\
\midrule
\begin{tabular}[c]{@{}c@{}}Num. of\\ hidden layer\end{tabular}                 & 7     & 4     & 8                 & 5                 \\
\midrule
\begin{tabular}[c]{@{}c@{}}Num. of\\ nodes in each\\ hidden layer\end{tabular} & 20    & 40    & 80                & 160               \\
\midrule
\begin{tabular}[c]{@{}c@{}}Num. of\\ epochs\end{tabular}                       & 300   & 300   & 300               & 300               \\
\midrule
Learning rate                                                                  & 0.001 & 0.001 & $5\times 10^{-5}$ & $5\times 10^{-5}$ \\ \bottomrule
\end{tabular}
\caption{\label{table:1} The hyperparameters and training details of the neural networks used in the NN-QREM experiments with two to five qubit in case of \texttt{ibmq\_belem} and \texttt{ibmq\_quito}.}
\end{table}

\begin{table}[ht]
\begin{tabular}{@{}>{\centering\arraybackslash}m{6.5em}|
>{\centering\arraybackslash}m{3.5em}
>{\centering\arraybackslash}m{3.5em}
>{\centering\arraybackslash}m{4.5em}
>{\centering\arraybackslash}m{4.5em}@{}}
\toprule
\begin{tabular}[c]{@{}c@{}}Num. of\\ qubits\end{tabular}                       & 2     & 3     & 4                 & 5                 \\ \midrule
\begin{tabular}[c]{@{}c@{}}Num. of\\ training data\end{tabular}                & 1800  & 3800  & 7800              & 9850              \\
\midrule
\begin{tabular}[c]{@{}c@{}}Num. of\\ test data\end{tabular}                    & 200   & 200   & 200               & 200               \\
\midrule
\begin{tabular}[c]{@{}c@{}}Num. of\\ hidden layer\end{tabular}                 & 5     & 2     & 7                 & 5                 \\
\midrule
\begin{tabular}[c]{@{}c@{}}Num. of\\ nodes in each\\ hidden layer\end{tabular} & 20    & 40    & 80                & 160               \\
\midrule
\begin{tabular}[c]{@{}c@{}}Num. of\\ epochs\end{tabular}                       & 300   & 300   & 300               & 300               \\
\midrule
Learning rate                                                                  & 0.001 & 0.001 & $5\times 10^{-5}$ & $5\times 10^{-5}$ \\ \bottomrule
\end{tabular}
\caption{\label{table:2} The hyperparameters and training details of the neural networks used in the NN-QREM experiments with two to five qubit in case of \texttt{ibmq\_qx2}.}
\end{table}

\subsection{Loss functions}
In this work, we evaluate three different distance measures (see Eq.~(\ref{eq:loss_function})) to quantify the amount of readout error mitigation and to compare the performance of different QREM methods. The metrics used for comparison are mean squared error (MSE) $$D_{\mathrm{MSE}}=\dfrac{1}{2^{n}}\sum\limits_{i=0}^{2^{n}-1}{|\bm{\tilde{p}}_i - \bm{p}_{i}|^{2}},$$
Kullback-Leibler divergence (KLD) $$D_{\mathrm{KLD}}=\sum\limits_{i=0}^{2^{n}-1}{\bm{p_{i}}\log{\dfrac{\bm{p}_i}{\bm{\tilde{p}}_i}}},$$ and
infidelity (IF)
$$D_{\mathrm{IF}}=1-\left(\sum\limits_{i=0}^{2^{n}-1}{\sqrt{\bm{p}_i\bm{\tilde{p}}_i}}\right)^2,$$
where $\bm{p}_i$ and $\bm{\tilde{p}}_i$ are the $i$th elements of the ideal and the mitigated probability distributions, respectively. For all these measures, a smaller value indicates a better performance as they represent the dissimilarity. Moreover, the infidelity is equivalent to the quantum state infidelity of two diagonal density matrices.

\subsection{Results}
\label{sec:results}

\begin{figure*}[t]
    \centering
    \includegraphics[width=0.99\textwidth]{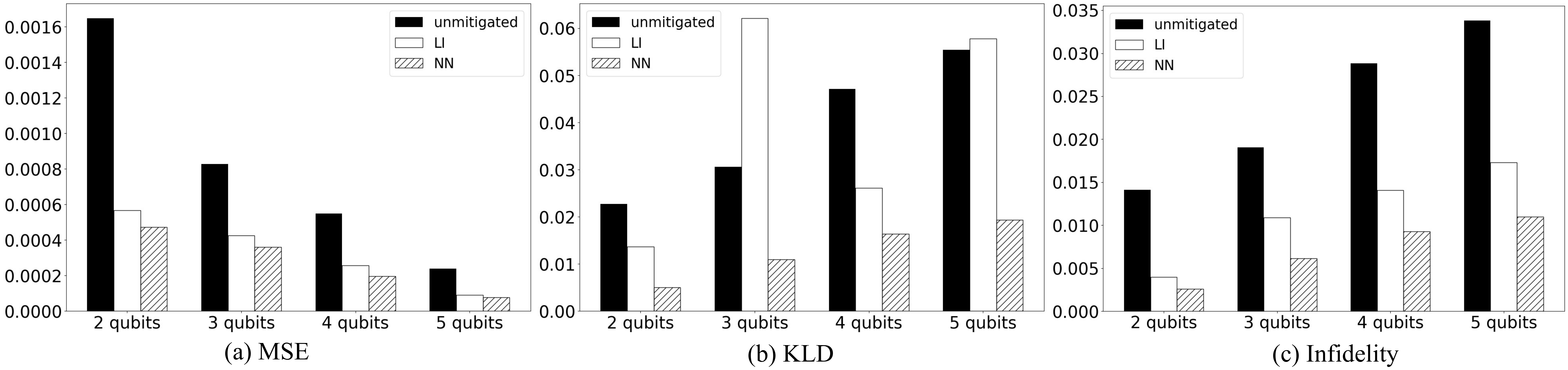}
    \caption{\label{fig:device_a_histogram} Experimental QREM results for reducing (a) MSE, (b) KLD and (c) infidelity from \texttt{ibmq\_quito} and \texttt{ibmq\_belem}. Filled bars represent unmitigated results, unfilled bars represent the LI-QREM results, and the hatched bars represent NN-QREM results.}
\end{figure*}
\vfill
\begin{figure*}[t]
    \centering
    \includegraphics[width=0.99\textwidth]{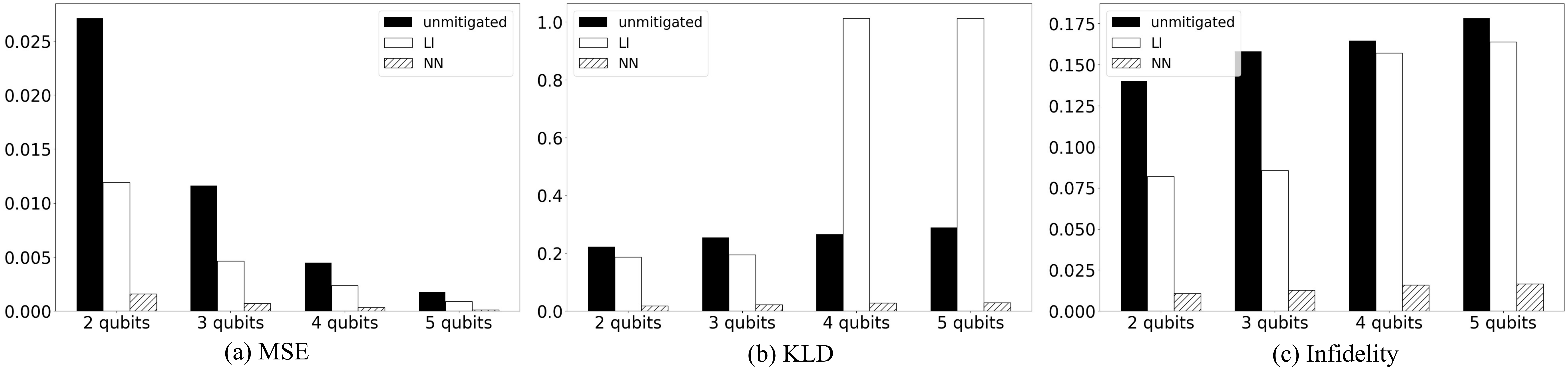}
    \caption{\label{fig:device_b_histogram} Experimental QREM results for reducing (a) MSE, (b) KLD and (c) infidelity from \texttt{ibmq\_qx2}. Filled bars represent unmitigated results, unfilled bars represent the LI-QREM results, and the hatched bars represent NN-QREM results.}
\end{figure*}
This section reports MSE, KLD and IF of (1) the raw (noisy) probability distribution $\bm{\hat{p}}$ (2) LI-QREM results $\bm{\tilde{p}}_\mathrm{LI}$ and (3) NN-QREM results $\bm{\tilde{p}}_\mathrm{NN}$ that are averaged over 200 test data. The QREM results are presented in Fig.~\ref{fig:device_a_histogram} and Fig.~\ref{fig:device_b_histogram} for the device type (a) and (b), respectively. The results clearly show that the NN-QREM can reduce the readout noise more effectively than the LI-QREM. Interestingly, LI-QREM is unable to reduce KLD in some cases, while NN-QREM reduces all quantifiers of the error in all cases.

To quantitatively compare the two methods, we define the performance improvement ratio for each loss function $D_{i}$ with the subscript $i$ being MSE, KLD, or IF, as
\begin{equation}
    R_i = \frac{D_{i}^{\mathrm{LI}} - D_{i}^{\mathrm{NN}}}{D_{i}^{\mathrm{NN}}}\times 100 (\%),
\end{equation}
where the superscript indicates whether the result is from LI-QREM or NN-QREM. By the definition of $R_i$, $R_i>0$ indicates that NN-QREM is better than LI-QREM, and vice versa. Table~\ref{table:3} shows the performance improvement ratio for all metrics. The values in the table show that NN-QREM outperforms LI-QREM in all instances. A notable observation is that NN-QREM works particularly better for the device type (b). We speculate that the readout noise in device (b) has higher non-linearities than that of device (a), albeit rigorous device-dependence analysis is left our for future work.

\begin{table}[ht]
\begin{tabular}{@{}c|
>{\centering\arraybackslash}m{4.5em}|
>{\centering\arraybackslash}m{3em}
>{\centering\arraybackslash}m{3em}
>{\centering\arraybackslash}m{3em}
>{\centering\arraybackslash}m{3em}@{}}
\toprule
 & \begin{tabular}[c]{@{}c@{}}Num. of\\ qubits\end{tabular} & 2 & 3 & 4 & 5 \\ \midrule
\multirow{3}{*}{\begin{tabular}[c]{@{}c@{}}Device\\ Type (a)\end{tabular}} & $R_{\mathrm{MSE}}$ & 19.9 & 18.4 & 30.4 & 17.6 \\
 & $R_{\mathrm{KLD}}$ & 174 & 466 & 59.8 & 200 \\
 & $R_{\mathrm{IF}}$ & 52.5 & 76.6 & 51.2 & 57.2 \\ \midrule
\multirow{3}{*}{\begin{tabular}[c]{@{}c@{}}Device\\ Type (b)\end{tabular}} & $R_{\mathrm{MSE}}$ & 648 & 564 & 611 & 592 \\
 & $R_{\mathrm{KLD}}$ & 888 & 775 & 3469 & 3305 \\
 & $R_{\mathrm{IF}}$ & 657 & 571 & 892 & 890  \\ \bottomrule
\end{tabular}
\caption{\label{table:3} Performance improvement ratio $R_i$ for all metrics tested in this work (i.e. MSE, KLD and IF) for each number of qubits. These numbers are categorized under two different device structures shown in Fig.~\ref{fig:devices}, labeled as Type (a) and Type (b)}
\end{table}

While the main purpose of our experiment is to demonstrate that the NN-based method is capable of mitigating the readout error beyond the existing linear inversion based methods, we point out that the neural networks used in this work can be further improved. For instance, one can choose to optimize the number of nodes in each hidden layer, instead of using some fixed number. One may also choose to use different activation functions and a different optimization algorithm, such as the Nesterov moment optimizer~\cite{Nesterov1983AMF}. One may also explore various loss functions for optimization. For example, one can choose the MSE, KLD or IF as the loss function to minimize through training, instead of the cross entropy loss used in this work.

Another interesting question is whether the neural network trained based on the data generated from one quantum device can be used to mitigate the noise of other quantum devices. This method is expected to work well especially for devices with the similar hardware characteristics. We tested this idea on \texttt{ibmq\_quito} and \texttt{ibmq\_belem}, since they have similar hardware design with the same qubit connectivity. In particular, we applied the neural network trained with the two- and three-qubit datasets generated by \texttt{ibmq\_quito} to mitigate the readout error in the dataset of \texttt{ibmq\_belem}. The corresponding QREM results are presented in Tab.~\ref{table:mix}. The experimental results show that the NN-QREM is better than the LI-QREM for all instances, except for reducing MSE in the case of three-qubit readout error.

\begin{table}[ht]
\begin{tabular}{@{}c|
>{\centering\arraybackslash}m{4.5em}|
>{\centering\arraybackslash}m{3em}
>{\centering\arraybackslash}m{3em}@{}}
\toprule
 & \begin{tabular}[c]{@{}c@{}}Num. of\\ qubits\end{tabular} & 2 & 3  \\ \midrule
\multirow{3}{*}{\begin{tabular}[c]{@{}c@{}}\texttt{ibmq\_quito} to \\ \texttt{ibmq\_belem} \end{tabular}} & $R_{\mathrm{MSE}}$ & 6.55 & -10.8 \\
 & $R_{\mathrm{KLD}}$ & 379 & 470  \\
 & $R_{\mathrm{IF}}$ & 82.5 & 71.4 \\ \bottomrule
\end{tabular}
\caption{\label{table:mix} Performance improvement ratio $R_i$ for three metrics (i.e. MSE, KLD and IF) for mitigating the readout error of \texttt{ibmq\_belem} using the datasets acquired from \texttt{ibmq\_quito}.}
\end{table}

\section{Conclusion}
\label{sec:conclusion}
This work proposes quantum readout error mitigation protocol based on deep learning with neural networks. The deep learning is known to be useful for finding non-linear relationships in a given dataset. This feature is utilized in our work to mitigate non-linear effects in the readout error. Hence our approach can achieve the level of error mitigation that is not attainable with the previous methods that rely on the linear error model. The advantage was clearly demonstrated through proof-of-principle experiments with two to five superconducting qubits. In all instances of experiments performed in this work, the neural network based QREM (NN-QREM) outperformed the linear inversion based QREM (LI-QREM) in reducing the error, which is quantified with three different measures: mean squared error, Kullback-Leibler divergence and infidelity. The improvement in error suppression is particularly more significant when Kullback-Leibler divergence and infidelity are considered. We also tested whether the data collected from one quantum device can be used to mitigate the readout error of the other devices. For example, we performed NN-QREM and LI-QREM to mitigate the readout error of \texttt{ibmq\_belem} using the data obtained from \texttt{ibmq\_quito}. Both methods can successfully mitigate the readout error when MSE and infidelity are considered, while only NN-QREM can also reduce KLD. Moreover, NN-QREM performed better than LI-QREM for this task in all instances, except in the case of three-qubits experiment with MSE.

Some remaining open problems and directions for future work are provided as follows. An important challenge in the domain of quantum error mitigation is the scalability of the algorithm, i.e. the total computational cost should not grow superpolynomially with the number of qubits. In our current construction, even if the number of training data is restricted to grow as $O(poly(n))$, the number of input and output nodes is $2^n$. However, the probability vector constructed by sampling from a quantum circuit is sparse as long as the number of repetition does not grow exponentially with $n$. Based on this observation, we plan to investigate machine learning methods for sparse data to explore the possibility of designing a scalable QREM algorithm. Another interesting future work is to develop a machine learning-based QREM protocol for the single-shot settings, which is important for many applications such as quantum teleportation, quantum error correction, and quantum communication.

\section*{Acknowledgment}
This research is supported by the National Research Foundation of Korea (Grant No. 2019R1I1A1A01050161 and Grant No. 2021M3H3A1038085), and Quantum Computing Development Program (Grant No. 2019M3E4A1080227). We acknowledge the use of IBM Quantum services for this work. The views expressed are those of the authors, and do not reflect the official policy or position of IBM or the IBM Quantum team.




\setcounter{equation}{0}
\setcounter{table}{0}
\setcounter{figure}{0}
\setcounter{section}{0}
\renewcommand{\theequation}{S\arabic{equation}}
\renewcommand{\figurename}{SUPPLEMENTARY FIG.}
\renewcommand{\tablename}{SUPPLEMENTARY TABLE}
\include{circuits}
\onecolumngrid
\bigskip
\begin{center}
{\bf Supplementary Information: Quantum readout error mitigation via deep learning}
\end{center}

\section{Optimization of the number of hidden layers}

This section explains the optimization procedure to decide the number of hidden layers of the neural network. 
In order to find the optimal number of hidden layers, $k$-fold cross-validation is carried out while changing the number of layers. The $k$-fold cross-validation process is summarized as follows:
\begin{enumerate}
    \item Set the neural network model and its hyperparameters.
    \item Split training data into $k$ sets of the same size.
    \item Select one set as a validation set, and the other sets as a training set.
    \item Train the neural network with the training set and calculate the score of the model through testing with the validation set.
    \item Iterate through steps 3 to 4 $k$ times and calculate the average score of the model.
    \item Iterate through steps 1 to 5 and select the model with the best average score.
\end{enumerate}

The work presented in the main manuscript selects $k$ to be 5 and uses infidelity (IF) as the score. After the process, the model with the smallest average infidelity is selected for QREM experiments. The 5-fold cross-validation method is depicted in Supplementary Fig.~\ref{fig:cross_validation}.

\begin{figure}[ht]
    \centering
    \includegraphics[width=0.7\columnwidth]{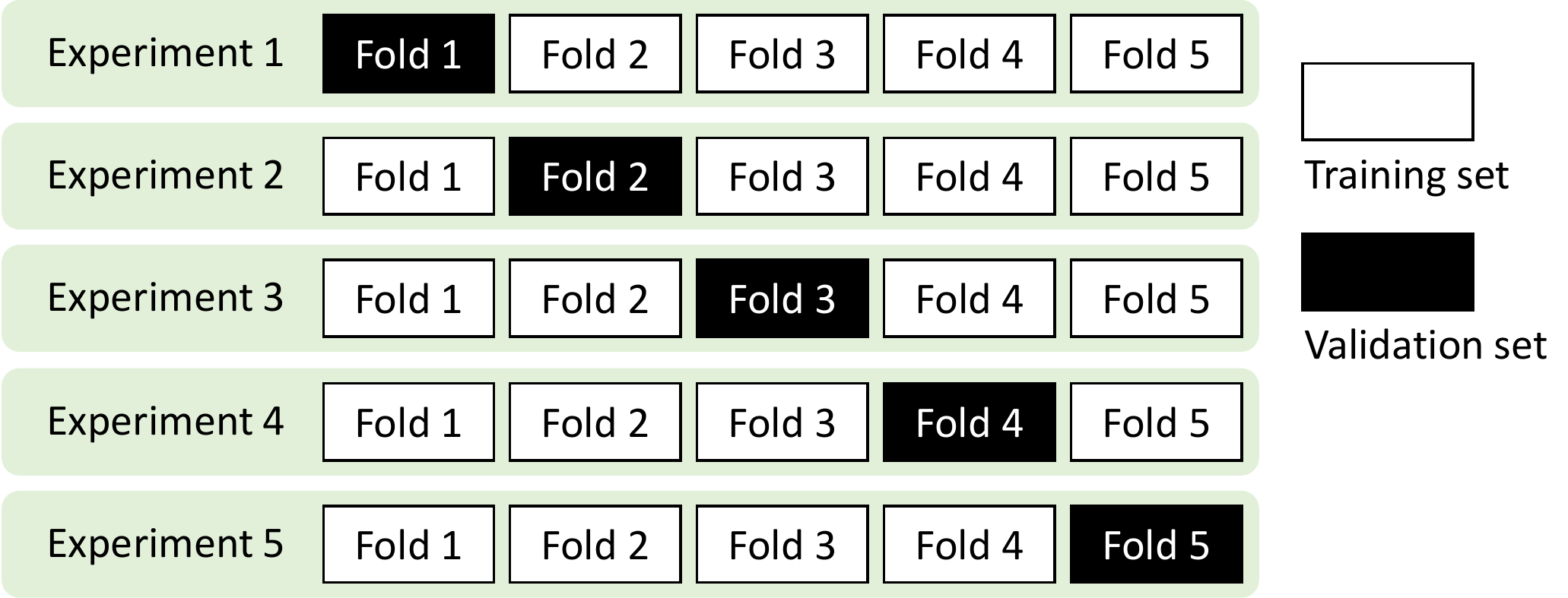}
    \caption{\label{fig:cross_validation} For 5-fold cross-validation, training set is divided into 5 subsets. The validation set is changed for each experiment and other sets that are not selected as the validation set remain as the training set. The black box indicates a validation set for the experiment. The white box indicates the training set for the experiment.
    }
\end{figure}

The average infidelity score as a function of the number of layers obtained during the 5-fold cross-validation process is shown in Supplementary Fig.~\ref{fig:line1} for quantum devices of type (a) and (b) (see Fig. 4 of the main manuscript for the device types).
\begin{figure}[ht]
    \centering
    \includegraphics[width=1\columnwidth]{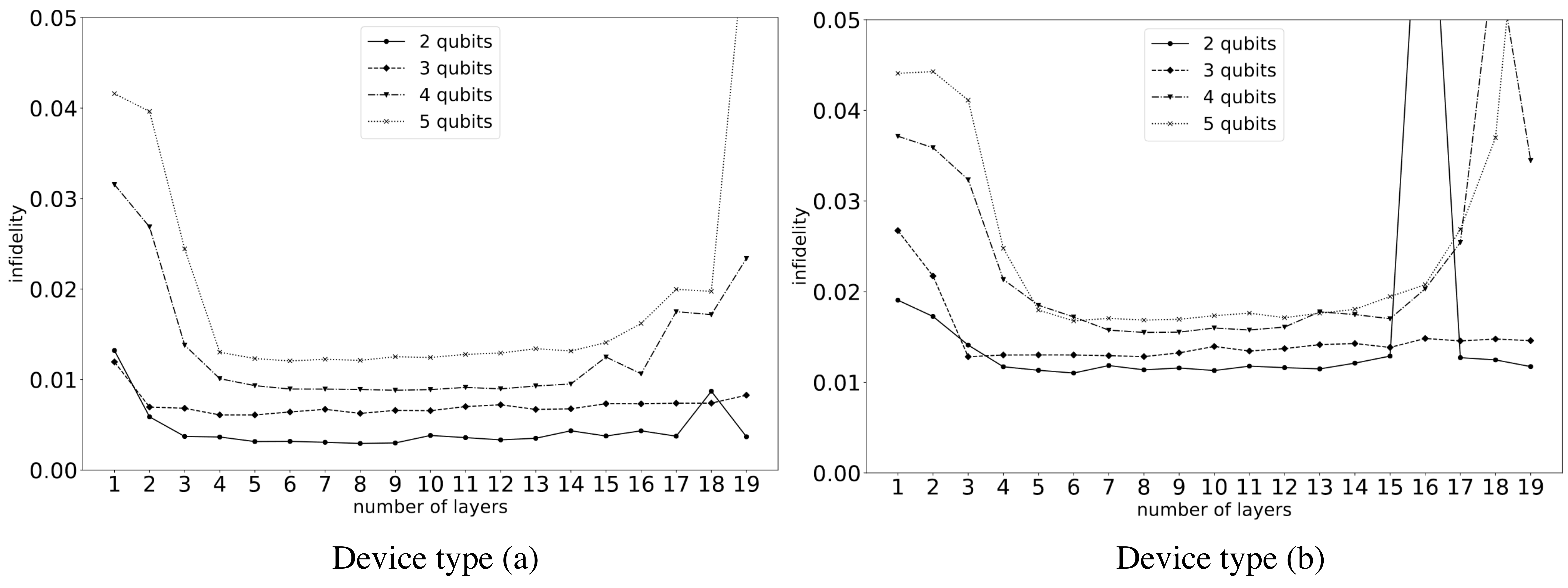}
    \caption{\label{fig:line1} The average infidelity as a function of the number of layers for the quantum device of type (a) and for the device of type (b). The number of qubits used in both device types ranges from two to five.}
\end{figure}

\section{Robustness of QREM methods against drift}

In general, the readout noise is not static. Hence QREM procedures described in the main manuscript needs to be carried out frequently as a part of the experimental calibration routine. Since QREM procedures can be costly, it is desirable to design QREM to be robust to the drift. 

\begin{figure}[h]
    \centering
    \includegraphics[width=0.9\columnwidth]{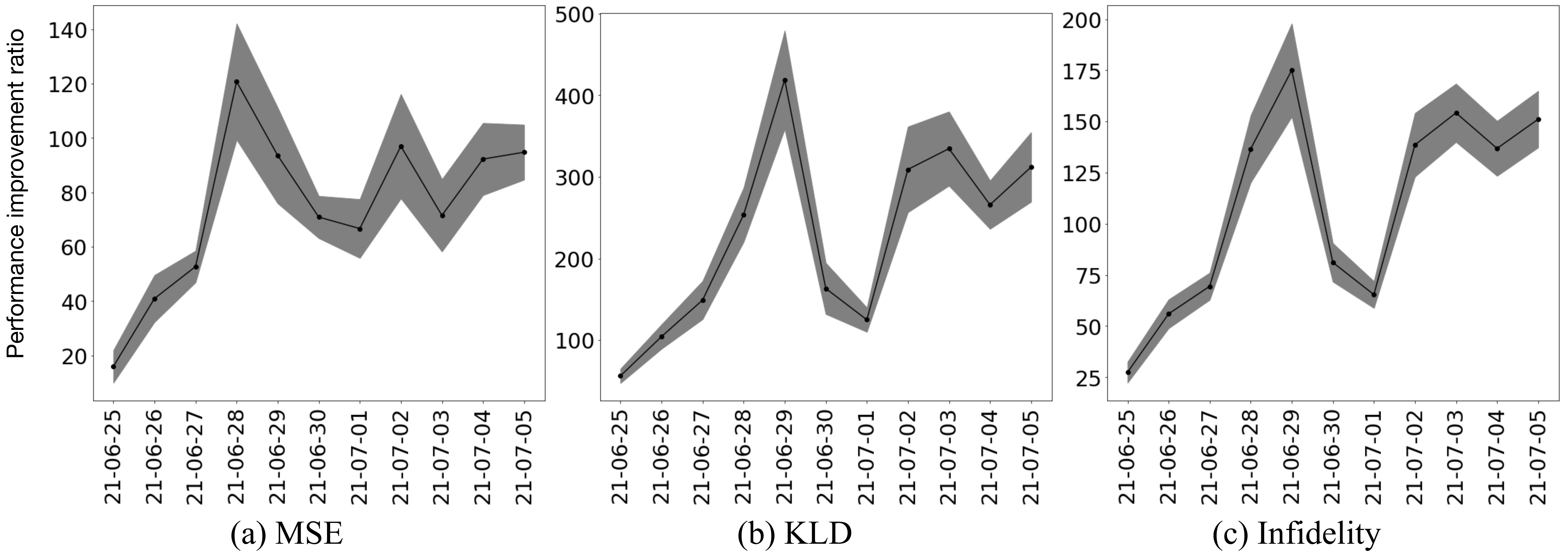}
    \caption{\label{fig:graph_under_time} The performance improvement ratio of (a) MSE, (b) KLD, and (c) IF obtained from 2021-06-25 to 2021-07-05 while the pre-trained neural network and the linear error matrix are fixed. The figure shows the mean values of 200 test data and the gray shaded region is the standard error. The training data for the neural network are collected from 2021-06-02 to 2021-06-11, and the data for the error matrix tomography are collected on 2021-06-17.}
\end{figure}

In the following, we show that the neural network-based QREM (NN-QREM) is more robust to the drift than the linear inversion-based QREM (LI-QREM). Recall that in the main manuscript the performance improvement ratio for each loss function $D_{i}$ with the subscript $i$ being MSE, KLD, or IF, is defined as
\begin{equation}
    R_i = \frac{D_{i}^{\mathrm{LI}} - D_{i}^{\mathrm{NN}}}{D_{i}^{\mathrm{NN}}}\times 100 (\%),
\end{equation}
where the superscript indicates whether the result is from LI-QREM or NN-QREM. By the definition of $R_i$, $R_i>0$ indicates that NN-QREM is better than LI-QREM, and vice versa. Supplementary Figure~\ref{fig:graph_under_time} shows the performance improvement ratio obtained with $\mathtt{ibmq\_belem}$ for 5-qubit QREM from 2021-06-25 to 2021-07-05 while the pre-trained neural network and the linear error matrix are fixed. The figure shows the mean values of 200 test data and the gray shaded region is the standard error. The training data for the neural network are collected from 2021-06-02 to 2021-06-11, and the data for the linear error matrix tomography are collected on 2021-06-17.

For all loss functions, $R_i$ is greater than zero at all times and it qualitatively increases over time, indicating that the NN-QREM is more robust to drift than LI-QREM.

\section{Training with smaller dataset}
The time required for training the neural network can be reduced by using a fewer number of training data. We tested the NN-QREM method with a smaller training dataset to see whether the good error mitigation performance can be retained while reducing the time cost. All hyperparameters and the test data are the same as the main paper, except the number of the training data is reduced by half. Despite having a smaller dataset, NN-QREM still performs better than LI-QREM as shown in SUPPLEMENTARY FIG. ~\ref{fig:reduced_bar_graph} and SUPPLEMENTARY TABLE ~\ref{table:reduced_per}. However, when two NN-QREM protocols with a different amount of training data are compared, the one with more data wins.

\begin{figure}[ht]
    \centering
    \includegraphics[width=1\columnwidth]{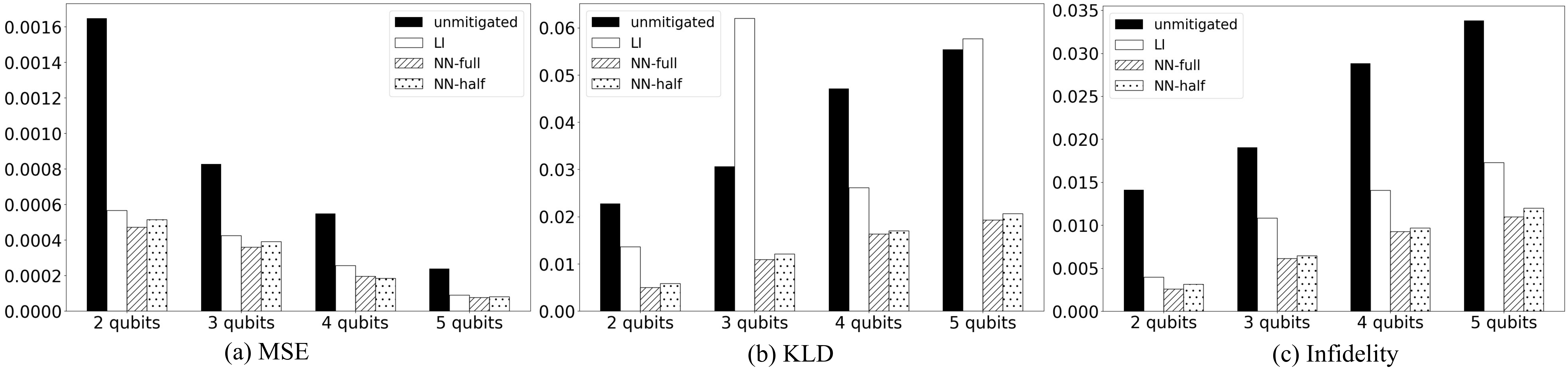}
    \caption{\label{fig:reduced_bar_graph}
    Experimental QREM results with two to five qubits for reducing (a) MSE, (b) KLD and (c) infidelity. Filled bars represent unmitigated results, unfilled bars represent the LI-QREM results, hatched bars represent the NN-QREM results with the full training dataset, and dotted bars represent the NN-QREM results results with the number of training dataset reduced by half.}
\end{figure}

\begin{table}[ht]
\begin{tabular}{@{}c
>{\centering\arraybackslash}m{4.5em}|
>{\centering\arraybackslash}m{3em}
>{\centering\arraybackslash}m{3em}
>{\centering\arraybackslash}m{3em}
>{\centering\arraybackslash}m{3em}@{}}
\toprule
 & \begin{tabular}[c]{@{}c@{}}Num. of\\ qubits\end{tabular} & 2 & 3 & 4 & 5 \\ \midrule
 & $R_{\mathrm{MSE}}$ & 9.77 & 8.79 & 38.0 & 12.8 \\
 & $R_{\mathrm{KLD}}$ & 133 & 411 & 53.5 & 179 \\
 & $R_{\mathrm{IF}}$ & 26.9 & 68.1 & 44.9 & 44.0 \\ \midrule
\end{tabular}
\caption{\label{table:reduced_per} Performance improvement ratio $R_i$ for all metrics tested in this work (i.e. MSE, KLD and IF) for each number of qubits. The number of training data for NN-QREM is a factor of two smaller than that used in the experiment presented in the main manuscript.}

\end{table}

\end{document}